# Integral Value Transformations: A Class of Affine Discrete Dynamical Systems and an Application

*Sk. S. Hassan[1], P. Pal Choudhury[1], B. K. Nayak[2], A. Ghosh[1] and J. Banerjee[1]*

*[1]Applied Statistics Unit, Indian Statistical Institute, Calcutta, India*

*[2]P. G. Department of Mathematics, Utkal University, Bhubaneswar, India*

*Email: sarimif@gmail.com, pabitrapalchoudhury@gmail.com, bknatuu@yahoo.co.uk, avishek.ghosh38@gmail.com and jogs.1989@gmail.com*

**_Abstract:_**

In this paper, the notion of *Integral Value Transformations* (IVTs), a class of Discrete Dynamical Maps has been introduced. Then notion of *Affine Discrete Dynamical System* (ADDS) in the light of IVTs is defined and some rudimentary mathematical properties of the system are depicted. *Collatz Conjecture* is one of the most enigmatic problems in 20th Century. The Conjecture was posed by German Mathematician *L. Collatz* in 1937. There are much advancement in generalizing and defining analogous conjectures, but even to the date, there is no prolific result for the advancement for the settlement of the conjecture. We have made an effort to make a Collatz type problem in the domain of IVTs and we have been able to solve the problem in 2011. Here mainly, we have focused and inquired on *Collatz-like ADDS*. Finally, we have designed the *Optimal Distributed and Parallel Environment* (ODPE) in the light of ADDS.

***Keywords:*** *Integral Value Transformations, Affine Discrete Dynamical System, Fractal, Collatz-like IVTs and Optimal Distributed Parallel Computing Environment.*

***1. Introduction:*** In [1, 2, and 3] the notion of *Integral Value Transformation* has been introduced and some rudimentary mathematics and discrete dynamical structures have been depicted. In this present study, the notion of *Affine Discrete Dynamical System* (ADDS) in the light of IVTs has been sketched. A relation has been established between the attractors of two type of ADDS and also the behavior of the attractor of the ADDS have been shown. An illustrious conjecture in Mathematics is *Collatz Conjecture*, posed by *L. Collatz* in 1937. The Collatz function T on $\mathbb{N}$ $to$ $\mathbb{N}$ is defined as $\mathrm{T(n) = 3n + 1}$; if n is odd, $\mathrm{T(n)} = \frac{\mathrm{n}}{2}$ ;if n is even.

The conjecture states that there exists a natural number $i$ such that the dynamical system $\mathrm{X_{n+1} = T(X_n)}$ carries any initial value $\mathrm{X_0}$ to $\mathrm{X_i} = 1$. In the similar fashion, ADDS in the light of IVTs has been presented in the next section. In the present study, a special emphasis has been given to those ADDS which are convergent and

1 *Corresponding Author's email: sarimif@gmail.com*

converges to a fixed point (attractor) for all the initial values. These ADDS named as *Collatz like ADDS*. The existence of such Collatz like ADDS has been confirmed earlier [1].

Finally, an *Optimal Distributed Parallel Computing Environment* (ODPE) has been designed through Collatz like ADDS.

## 2. Collatz like IVTs and Discrete Dynamical System

### 2.1 *Notion of Integral Value Transformations (IVTs):*

Integral Value Transformations (IVTs) is a class of continuous maps in a discrete space $\mathbb{N}_0 = \mathbb{N} \cup \{0\}$.

***Definition 2.1*** A p-adic, k-dimensional, Integral Value Transformation is denoted by $IVT^{p,k}{}_j$. $IVT^{p,k}{}_j$ from ${\mathbb{N}_0}^K$ to $\mathbb{N}_0$ is defined as

$$IVT^{p,k}{}_j((n_1, n_2, \dots . n_k)=$$

$$(f_j({a_0}^{n_1}, {a_0}^{n_2}, \dots, {a_0}^{n_k})\, f_j({a_1}^{n_1}, {a_1}^{n_2}, \dots, {a_1}^{n_k}) \dots \dots f_j({a_{l-1}}^{n_1}, {a_{l-1}}^{n_2}, \dots, {a_{l-1}}^{n_k}))_p = m$$

where $n_1 = ({a_0}^{n_1}{a_1}^{n_1} \dots . {a_{l-1}}^{n_1})_p$, $n_2 = ({a_0}^{n_2}{a_1}^{n_2} \dots . {a_{l-1}}^{n_2})_p, \dots \dots$ $n_k = ({a_0}^{n_k}{a_1}^{n_k} \dots . {a_{l-1}}^{n_k})_p$

$f_j: \{0, 1, 2, \dots, p-1\}^k \rightarrow \{0,1,2, \dots, p-1\}$ and m is the decimal conversion from the p adic number.

Let us fix the domain of IVTs as $\mathbb{N}_0$ (k=1) and thus the above definition boils down to the following:

$$IVT^{p,1}{}_j\,(x) = \left(f_j(x_n)\, f_j(x_{n-1}) \dots \dots \dots f_j(x_1)\right)_p = m$$

where m is the decimal conversion from the p adic number, and $x = (x_1\ x_2 \dots \dots x_n)_p$.

Now, let us denote the set of $IVT^{p,1}{}_j$ as

$$T^{p,1} = \left\{ IVT^{p,1}{}_j \ : \mathbb{N}_0 \rightarrow \mathbb{N}_0 \ \middle|\ \begin{array}{c} 0 \le j < p^p,\ IVT^{p,1}{}_j\,(x) = \left(f_j(x_n)\, f_j(x_{n-1}) \dots \dots \dots f_j(x_1)\right)_p = m \\ \text{where m is the decimal conversion from the p adic number} \\ \text{and } x = (x_1\ x_2 \dots \dots x_n)_p \end{array} \right\}$$

The definition is illustrated below:

For p=3, k=1

| **p=3** | $\mathbf{f_7}$ | $\mathbf{f_{16}}$ |
|---|---|---|
| **0** | 1 | 1 |
| **1** | 2 | 2 |
| **2** | 0 | 1 |

$x = 55 = (\,2001\,)_3$

$IVT^{3,1}{}_7(x) = (\,f_7(2)\ f_7(0)f_7(0)f_7(1))_3 = (\,0112)_3 = 14$

& $IVT^{3,1}{}_{16}(x) = \left(\,f_{16}(2)f_{16}(0)f_{16}(0)f_{16}(1)\right)_3 = (\,1112\,)_3 = 41$

***Definition 2.2:*** A function f is said to be a *Collatz-like function* if $\exists\ m \in \mathbb{N}_0$ such that $f^m(n) = c$, c is a fixed point over the iterations for any choice of n.

Some IVTs, defined as above, are seen to converge to a fixed point c (say) after finitely many iterations i.e. $\exists\ m \in \mathbb{N}_0$ such that $(\mathrm{IVT}^{p,1}{}_j)^m(n) = c$ for any choice of n.

Thus the iterative scheme $X_{n+1} = \mathrm{IVT}^{p,1}{}_j\,(X_n), n = 0, 1, \ldots n, \ldots\ldots$ converges to the fixed point c for any given $X_0$ (belongs to $\mathbb{N}_0$). Such functions are called *Collatz-like Integral Value Transformations.*

In any p-adic system, there are $(p^{p-1} - 1)$ number of Collatz-like IVTs [1].

***2.2 Discrete Dynamical System***

***Definition 2.3***: A semi-group $(G, f)$ where $f\colon G \times G \to G$ is the binary operation, acting on a space M is called a dynamical system if a mapping

T: G x M → M defined as T (g, x) = $T_g$(x) such that $T_{f(g,h)} = f(T_g, T_h)$. If G is a discrete set, then the system is called a *Discrete Dynamical System* [*DDS*] [4]. Further, if $G = \mathbb{N}_0$ or G = Z, IVTs form a *discrete dynamical system* when applied iteratively and this opens up a vast unexplored area. It is interesting to see how these functions evolve over time, form chaotic patterns, etc. The real motivation is to make an attempt at understanding how these IVTs evolve over time. Dynamical systems could throw some light in this respect thereby aiding us in comprehending the time evolution of these functions.

Therefore the iterative scheme $X_{n+1} = \mathrm{IVT}^{p,1}{}_j\,(X_n), n = 0, 1, \ldots n, \ldots\ldots$ can be thought as a one dimensional non-linear discrete dynamical system [3, 4].

***Definition 2.4***: A steady state equilibrium of the equation $x_n = \mathrm{IVT}^{p,1}{}_j(x_{n-1})$ is a point $\bar{x} \in \mathbb{N}_0$ such that $\mathrm{IVT}^{p,1}{}_j(\bar{x}) = \bar{x}$ that is $\bar{x}$ is a fixed point of $\mathrm{IVT}^{p,1}{}_j$ .

Stability Analysis of steady state equilibria of discrete dynamical systems is based on some propositions and/or explicit solution of the non-linear, autonomous (the parameters / coefficients *a* and *b* in the difference equation $x_n = ax_{n-1} + b$ are independent of time), one-dimensional dynamical systems after reducing the non-linear system to a linear system [4].

A linear system is called ***locally stable*** if for a small perturbation to the system, it converges asymptotically to the original equilibrium. A linear system is called ***globally stable*** if irrespective of the extent of perturbation, it converges asymptotically to the original equilibrium. Mathematically, the definition is as follows:

***Definition 2.5*:** A steady state equilibrium $\bar{x}$, of the linear difference equation $x_n = a x_{n-1} + b$ is called ***globally (asymptotically) stable*** if $\lim_{n\to\infty} x_n = \bar{x}$ for all $x_0 \in \mathbb{N}_0$ and is called ***locally (asymptotically) stable*** if there exists $r > 0$ such that $\lim_{n\to\infty} x_n = \bar{x}$ for all $x_0 \in N(\bar{x}, r)$, a neighbourhood of the point $\bar{x}$.

Now, for the non-linear dynamical system $x_n = \mathrm{IVT}^{p,1}{}_j(x_{n-1})$ …… (1)

Through the Taylor series expansion of the non-linear system around the fixed point $\bar{x}$ of (1), it can be reduced to a linear system around the steady state equilibrium $\bar{x}$ to approximate the behaviour around the fixed point by a linear system. The Taylor series expansion is the following

$$x_{n+1} = \mathrm{IVT}^{p,1}{}_j(x_n) = \mathrm{IVT}^{p,1}{}_j(\bar{x}) + (x_n - \bar{x})\mathrm{DIVT}^{p,1}{}_j(\bar{x}) + \frac{(x_n - \bar{x})^2}{2!}$$

$D^2 IVT^{p,1}{}_j(\bar{x}) + \ldots\ldots\ldots + \frac{(x_n-\bar{x})^k}{k!} D^k IVT^{p,1}{}_j(\bar{x}) + \ldots\ldots$ where $D^k IVT^{p,1}{}_j(\bar{x})$ denotes the kth derivative of $IVT^{p,1}{}_j$ at the point $\bar{x}$. The concept of derivative in $T^{p,1}$ is defined in [3]

The linearized system around the fixed point is

$x_{n+1} = IVT^{p,1}{}_j(\bar{x}) + (x_n - \bar{x}) DIVT^{p,1}{}_j(\bar{x})$ neglecting the higher order terms

= $[DIVT^{p,1}{}_j(\bar{x})] x_n + [IVT^{p,1}{}_j(\bar{x}) - \bar{x} DIVT^{p,1}{}_j(\bar{x})]$ = $ax_{n-1} + b$ ……..(2)

where $a = DIVT^{p,1}{}_j(\bar{x})$ and $b = [IVT^{p,1}{}_j(\bar{x}) - \bar{x} DIVT^{p,1}{}_j(\bar{x})]$

By a proposition stated below

The linearized system (2) is globally stable iff $|a| = |DIVT^{p,1}{}_j(\bar{x})| < 1$

That is, for any $x_0 \neq \bar{x}, \lim_{n\to\infty} x_n = \bar{x}$ if $|a| = |DIVT^{p,1}{}_j(\bar{x})| < 1$ and convergence is monotonic if 0<a<1 and oscillatory if $-1 < a < 0$.

So far, we have seen the stability condition of the dynamical system (1). Now, we are about to investigate Affine Discrete Dynamical System (ADDS).

## 3. *Affine Discrete Dynamical System (ADS) and their characterization:*

### *3.1 Definitions of ADDS*

Let us first define what we mean by *Affine Discrete Dynamical System* (ADDS).

**Definition 3.1:** The *ADDS of Type-I* is defined as

$$Y_{t+1} = A\, IVT^{p,1}{}_j\,(Y_t) + B;\ A = 1, 2, \ldots., p-1.\ B = 0, 1, 2, .., p-1. \ldots\ldots (2)$$

With $y_0$ as the initial state with $t = 0$.

Hence,

$$Y_1 = A\, IVT^{p,1}{}_j\,(Y_0) + Y_2 = A\, IVT^{p,1}{}_j\,(Y_1) + B = A\, IVT^{p,1}{}_j\,\big(A\, IVT^{p,1}{}_j\,(Y_0) + B\big) + B,$$

Similarly we have $y_3 = A\, IVT^{p,1}{}_j\,\big(A\, IVT^{p,1}{}_j\,\big(A\, IVT^{p,1}{}_j\,(Y_0) + B\big) + B\big) + B$

Hence, the generalized form of the above iterative definition can be summarized as

$$y_{\,t} = A\, IVT^{p,1}{}_j\,(A\, IVT^{p,1}{}_j\,(A\, IVT^{p,1}{}_j\,(\ldots\ldots. IVT^{p,1}{}_j\,(A\, IVT^{p,1}{}_j\,(Y_0) + B) + \ldots\ldots.. + B) + B$$

**Definition 3.2:** The *ADDS of Type-II* is defined as

$y_{t+1} = IVT^{p,1}{}_j\,(a\, y_{\,t} + b\,);\ a = 1,2, \ldots., p-1.\ b = 0,1,2, .., p-1$ …… (3)

With $y_0$ as the initial state with $t = 0$ Hence,

$$y_{\,1} = IVT^{p,1}{}_j\,(a\, y_0 + b), y_{\,2} = IVT^{p,1}{}_j\,(a\, y_1 + b) = IVT^{p,1}{}_j\,(a\, IVT^{p,1}{}_j\,(a\, y_0 + b) + b)\,,$$

Similarly, we have $y_{\,3} = IVT^{p,1}{}_j\,(\,a\, IVT^{p,1}{}_j\,(a\, IVT^{p,1}{}_j\,(a\, y_0 + b) + b) + b)$

Hence, the generalized form of the above iterative definition can be summarized as

$$y_t = IVT^{p,1}{}_j\,(a\, IVT^{p,1}{}_j\,(a\, IVT^{p,1}{}_j\,(\ldots\ldots. IVT^{p,1}{}_j\,(a\, IVT^{p,1}{}_j\,(a\, y_0 + b) + b) + \ldots\ldots.. + b)$$

Let us now investigate the stability condition of the above to type ADDS viz. $X_{t+1} = a\, f\,(X_t) + b$ and $X_{t+1} = f\,(a\, X_{\,t} + b\,)$ .

### *3.2 Steady State Equilibrium:*

**Definition:** A system is said to reach a steady state equilibrium iff, once the state is reached the system will remain in that state for future iterations [4].

i.e. $Y_{t+1} = Y_t$ Also, .... = $Y_{t+2} = Y_{t+1} = Y_t$ (successive iterative value after t remains the same). For the above definition of affine class, let ȳ be the steady state equilibrium. Then the ADDS of type-I and ADDS of type-II become $\bar{y} = A\,\mathrm{IVT}^{p,1}{}_j(\bar{y}) + B$ and $\bar{y} = \mathrm{IVT}^{p,1}{}_j(a\bar{y} + b)$ respectively.

Solution of the above equation will yield steady state equilibrium points. It may or may not be unique depending upon $\mathrm{IVT}^{p,1}{}_j, a, b$.

To study **local stability** of the ADDS we need to make the iterative scheme (2) and (3) into a Linearized system. $\mathrm{IVT}^{p,1}{}_j(Y_t)$ can be expanded about ȳ by Taylor series as

$$\mathrm{IVT}^{p,1}{}_j(Y_t) = \mathrm{IVT}^{p,1}{}_j(Y') + \mathrm{DIVT}^{p,1}{}_j(\bar{y})(Y_t - \bar{y}) + (\mathrm{D}^2\mathrm{IVT}^{p,1}{}_j(\bar{y})(Y_t - \bar{y})2)/2! + \dots\dots..$$

For linearization purpose and considering a small neighborhood around ȳ, first 2 terms will be taken into account

$\mathrm{IVT}^{p,1}{}_j(Y_t) = \mathrm{IVT}^{p,1}{}_j(\bar{y}) + \mathrm{DIVT}^{p,1}{}_j(\bar{y})(Y_t - \bar{y})$

Substituting the approximated value to the governing dynamical equations we get

$$Y_{t+1} = A\,\mathrm{IVT}^{p,1}{}_j\left(\mathrm{IVT}^{p,1}{}_j(\bar{y}) + \mathrm{DIVT}^{p,1}{}_j(\bar{y})(Y_t - \bar{y})\right) + B$$

$$= A\,\mathrm{DIVT}^{p,1}{}_j(\bar{y})Y_t + A\,\mathrm{IVT}^{p,1}{}_j(\bar{y}) + B - \mathrm{DIVT}^{p,1}{}_j(\bar{y})\bar{y} = \mathcal{A}Y_t + \mathcal{B}$$

Where, $\mathcal{A} = A\,\mathrm{DIVT}^{p,1}{}_j(\bar{y})$ & $\mathcal{B} = A\,\mathrm{IVT}^{p,1}{}_j(\bar{y}) + B - \mathrm{DIVT}^{p,1}{}_j(\bar{y})\bar{y}$.

After linearization, the system behaves like a linear system with $Y_{t+1} = \mathcal{A}\,Y_t + \mathcal{B}$, in small neighborhood of ȳ. The condition of local stability is enumerated below $|\mathcal{A}| < 1$

- $|A\,D\mathrm{IVT}^{p,1}{}_j(\bar{y})| < 1$
- $|D\mathrm{IVT}^{p,1}{}_j{}'(\bar{y})| < 1/A$ ( As $A$ is positive )

Similarly, $\mathrm{IVT}^{p,1}{}_j(a\,y_t + b)$ can be expanded about ȳ by Taylor series. For linearization purpose and considering a small neighborhood around ȳ, first 2 terms will be taken into account,

$$\mathrm{IVT}^{p,1}{}_j(a\,y_t + b) = \mathrm{IVT}^{p,1}{}_j(a\bar{y} + b) + a\,\mathrm{DIVT}^{p,1}{}_j(a\bar{y} + b)(y_t - \bar{y}) + \dots\dots..$$

So the governing dynamical equation is

$$y_{t+1} = \mathrm{IVT}^{p,1}{}_j(a\,y_t + b) = \mathrm{IVT}^{p,1}{}_j(a\bar{y} + b) + a\,\mathrm{DIVT}^{p,1}{}_j(a\bar{y} + b)(y_t - \bar{y})$$

$$= a\,\mathrm{DIVT}^{p,1}{}_j(a\bar{y} + b)\,y_t + \mathrm{IVT}^{p,1}{}_j(a\bar{y} + b) - a\,f'(a\bar{y} + b)\,\bar{y} = \mathcal{A}\,y_t + \mathcal{B}$$

Where, $\mathcal{A} = a\,\mathrm{DIVT}^{p,1}{}_j(a\bar{y} + b)$ & $\mathcal{B} = \mathrm{IVT}^{p,1}{}_j(a\bar{y} + b) - a\,\mathrm{DIVT}^{p,1}{}_j(a\bar{y} + b)\bar{y}$

After linearization, the system behaves like a linear system with $y_{t+1} = \mathcal{A}\, y_t + \mathcal{B}$, in small neighborhood of $\bar{y}$.

The condition of local stability is given below

$$|\mathcal{A}| < 1$$

$$\Rightarrow \left| a\, \mathrm{DIVT}^{p,1}{}_{j}\,(a\ \bar{y} + b)\right| < 1 \Rightarrow |\,\mathrm{DIVT}^{p,1}{}_{j}(a\,\bar{y} + b)\,| < 1/a \quad (\text{As } a \text{ is positive})$$

In order to achieve **global stability** of the ADDS, the $\mathrm{IVT}^{p,1}{}_{j}$ has to be a contraction or Lipchitz function. The system $Y_{t+1} = A\, \mathrm{IVT}^{p,1}{}_{j}\,(Y_t) + B$, is *globally stable* if

$$\frac{\left|\left(A\ \mathrm{IVT}^{p,1}{}_{j}\,(Y_{t+1}) + B\right) - \left(A\ \mathrm{DIVT}^{p,1}{}_{j}\,(Y_t) + B\right)\right|}{|Y_{t+1} - Y_t|} < 1\ \forall\, t = 0,1,2,\ldots\ldots,\infty$$

$$\Rightarrow |A\mathrm{DIVT}^{p,1}{}_{j}(Y_t)| < 1 \quad \forall\, t = 0,1,2,\ldots\ldots,\infty$$

If the dynamical system shows contraction, then only it has *globally stable* points.

For ADDS type-II to be globally stable this condition becomes

$|a\mathrm{DIVT}^{p,1}{}_{j}(ay_t + b)| < 1 \quad \forall\, t = 0,1,2,\ldots\ldots,\infty$

## *4. Stability Analysis of ADDS:*

### 4.1. Illustration of Local Stability:

Let us first deal with ADDS of type-I, $Y_{t+1} = A\ \mathrm{IVT}^{2,1}{}_{j}\,(Y_t) + B$ where A, B are either 0 or 1.

There are only four IVTs in $\mathrm{T}^{2,1}$ namely $\mathrm{IVT}^{2,1}{}_{0}, \mathrm{IVT}^{2,1}{}_{1}, \mathrm{IVT}^{2,1}{}_{2}$ & $\mathrm{IVT}^{2,1}{}_{3}$.

In case of $\mathrm{IVT}^{p,1}{}_{0}\,(x) = 0$ (Zero function)

Steady state equilibrium points $Y_{t+1} = Y_t = \bar{y}$

$$\bar{y} = A\ \mathrm{IVT}^{p,1}{}_{0}\,(\bar{y}) + B \Rightarrow \bar{y} = B \quad (As\ \mathrm{IVT}^{p,1}{}_{0}(x) = 0)$$

It is unique as it is independent of the initial point $y_0$. (i.e.unique.)

Also $\mathrm{DIVT}_0^{2,1}(\mathrm{x}) = 0$

$$So, \mathrm{DIVT}^{2,1}{}_{0}\,(\bar{y}) = 0$$

$$So, |\,A\,D\mathrm{IVT}^{2,1}{}_{0}\,(\bar{y})\,| = 0 < 1$$

So at $\bar{y} = B$ points are locally stable. Also at $\bar{y} = B$ shows global stability as $|AD\mathrm{IVT}^{2,1}{}_{0}(y_t)| = 0 < 1 \quad \forall\, t = 0,1,2,\ldots\ldots,\infty$

In case of $\mathrm{IVT}^{2,1}{}_{1} = (2^k - 1) - x$ where $x$ is of $k$ bit.

Steady state equilibrium points $\bar{y} = A\ \mathrm{IVT}^{2,1}{}_{1}(\bar{y}) + B = A(2^k - 1 - \bar{y}) + B$ Where $\bar{y}$ is represented in $k$ bits.

$\Rightarrow \bar{y}(1 + A) = A(2^k - 1) + B \Rightarrow \bar{y} = \frac{A(2^k-1)+B}{(1+A)}$ ………. (4)

Unfortunately, nothing can be concluded from equation (4) about the existence of steady state points. However, from its dynamical behavior conclusion can be drawn. The function is Collatz like only if $A = 1, B = 0$. In that case also the function oscillates around its attractor $(0)$. No steady state solution exists for this case. Other combination of $A$ and $B(A \neq 0)$ makes the function non-Collatz yielding no steady state points.

The function is not differentiable (i.e.$D$ [$\mathrm{IVT}_1^{2,1}$ $(x)$ ] does not exist). So no question of stability (local or global) arises for this function. Only $A = 0$ leads to a constant system having unique steady state points and thus locally and globally stable solutions.

In case of $\mathrm{IVT}^{2,1}{}_2(x) = x$, (Identity function)

Steady state equilibrium $Y_{t+1} = Y_t \Rightarrow \bar{y} = A\,\bar{y} + B$

$\Rightarrow \bar{y}\ \left(1 - A\right) = B \Rightarrow \bar{y} = B\,/\,(1 - A)\ \ldots.(5)$

Now in order to get Steady state equilibrium, $\bar{y}$ must be positive integer (including $0$ ) and $A, B$ are to be non-negative integers ($0 \leq A, B \leq 1$).

Considering the constraints equation (5) has the following solutions

(i) If $A = 0,\ \bar{y} = B$ ( Leading to a constant system as in case of $\mathrm{IVT}_0^{2,1}$)

(ii) If $A = 1, B \neq 0, \bar{y} \to \infty$ i.e. the system has equilibrium point at infinity which is practically not feasible).

(iii) if $A \neq 1, B = 0,\ \bar{y} = 0$ ( $0$ is the solution )

(iv) If $A = 1, B = 0,\ \bar{y} = y_0$(Non unique steady state solution)

It is obvious that $\mathrm{DIVT}_2^{2,1}$ (x) $= 1$ So, $\mathrm{DIVT}_2^{2,1}\ (\bar{y}) = 1$

Condition for local stability, $|\ A\ \mathrm{DIVT}_2^{2,1}\ (\bar{y})| < 1$. i.e. $|\ A\ | < 1$ and this is possible only if $A = 0$ which leads to a constant system therefore the local stability attains at $Y_{t+1} = B$ as depicted for as $A$ is $\mathrm{IVT}_0^{2,1}$.

$\mathrm{IVT}_3^{2,1}(x) = 2^k - 1$ where $x$ is of $k$ bit.

Steady state equilibrium points $\bar{y} = A\ \mathrm{IVT}_3^{2,1}(\bar{y}) + B = A(2^l - 1) + B$

Where $\bar{y}$ is represented in $l$ bits.

The above IVT is not Collatz like, not a differentiable and so none of the combinations of $A$ and $B$ (unless $A = 0$) will yield steady-state solutions and consequently leading to the stability of the dynamical system.

Let us now see the same for ADDS of type-II $y_{t+1} = \mathrm{IVT}_j^{2,1}\ (a\ y_{\ t} +\ b\ )$as the following:

In case of $\mathrm{IVT}_0^{2,1}$, $\mathrm{IVT}_0^{2,1}\ (x) = 0$, (Trivial function)

Steady state equilibrium points $y_{t+1} = y_t = \bar{y}$

$$\bar{y} = \mathrm{IVT}_0^{2,1}\ (a\,\bar{y} + b\,) \Rightarrow \ \bar{y} = 0 \quad (As\ \ \mathrm{IVT}_0^{2,1}(x) = 0)$$

It is unique as it is independent of the initial point $y_0$.

Also $\mathrm{DIVT}_0^{2,1}$(x) $= 0\ So,\ \mathrm{DIVT}_0^{2,1}\ (a\,\bar{y} + b) = 0$

$$So, |\ a\ \ \mathrm{IVT}_0^{2,1}\ (\,a\bar{y} + b)\ | = 0 < 1$$

$$\text{Also, } |a\ \mathrm{DIVT}_0^{2,1}(ay_t\text{+b})| < 1 \quad \forall\, t = 0,1,2, \ldots\ldots, \infty$$

Therefore the solution $\bar{y} = 0$ is unique, locally and globally stable.

In case of $\mathrm{IVT}_1^{2,1}(x) = (2^k - 1) - x$ where $x$ has $k$ bit representation.

Steady state equilibrium points $\bar{y} = \mathrm{IVT}_1^{2,1}(a\,\bar{y} + b) = (2^k - 1 - a\,\bar{y} - b)$ Where $a\bar{y} + b$ is represented in $k$ bits.

As explained earlier nothing can be concluded from the above equation. From dynamical behavior the function is Collatz like only if a=1, b=0 and in that case also it oscillates around attractor (0). Also it is a non-differentiable function. So no steady state solution exists, no question of local or global stability.

In case of $\mathrm{IVT}_2^{2,1}(x) = x$, (Identity function)

For Steady state equilibrium $y_{t+1} = y_t \Rightarrow \bar{y} = a\,\bar{y} + b \Rightarrow \bar{y}\,(1 - a) = b \Rightarrow \bar{y} = b\,/\,(1 - a)$

The conclusions are same as in ADDS of type I. In case of $\mathrm{IVT}_3^{2,1}(x) = 2^k - 1$ where $x$ $k$ bit representation.

Steady state equilibrium points $\bar{y} = \mathrm{IVT}_3^{2,1}(a\,\bar{y} + b) = (2^l - 1)$ where $a\bar{y} + b$ is represented in $l$ bits.

The function is non-Collatz like and increasing function. As explained for $\mathrm{IVT}_3^{2,1}$ in ADDS of type I.

**4.2 Illustration of Global Stability:**

Let us explore the global stability condition of ADDS of type-I and type-II in 2-adic system as we did in previous section.

Contraction condition of the iterative function is the primordial need to be globally stable of a dynamical system. The condition of contraction is stated below as a definition.

**Definition 4.2.1:** A function $f$ is said to be a *contraction* if and only if $f: X \to X$ ($X$ is the vector space) satisfies $d\big(f(x), f(y)\big) \leq \lambda\, d(x,y)\ \forall\, x, y \in X$ where $\lambda \in (0,1)$ and $d$ is the metric defined on the vector space.

In present case the underlying space is $\mathbb{N}_0$ and usual Euclidian metric, $d(x,y) = |x - y|$ can be employed.

We have studied for $(A = 1, B = 0)$ in 2-adic linearized system.

For the function, $\mathrm{IVT}_0^{2,1}(x) = 0$,

The contraction condition becomes $\frac{|\mathrm{IVT}_0^{2,1}(x) - \mathrm{IVT}_0^{2,1}(y)|}{|x-y|} = 0 < \lambda\,;\lambda \in (0,1)$

Therefore $\mathrm{IVT}_0^{2,1}$ is a contraction.

For the function $\mathrm{IVT}_1^{2,1}(x) = (2^k - 1) - x$ where $x$ is of $k$ bit number.

Now,

$\frac{d(\mathrm{IVT}_1^{2,1}(x), \mathrm{IVT}_1^{2,1}(y))}{d(x,y)} = \left|\frac{(2^k - x) - (2^l - y)}{(x-y)}\right|$ where $x$ and $y$ has $k$ and $l$ bit representation respectively.

Consider the case where x and y has same number of bits i.e. k=l, so $\frac{d(\mathrm{IVT}_1^{2,1}(x), \mathrm{IVT}_1^{2,1}(y))}{d(x,y)} = 1 \nleq \lambda$; $\lambda \in (0,1)$.

So $d\left(\mathrm{IVT}_1^{2,1}(x), \mathrm{IVT}_1^{2,1}(y)\right) \nleq \lambda\, d(x,y)\ \forall\, x, y \in X$. Thus$\mathrm{IVT}_1^{2,1}$ is not a contraction.

It is trivial that $\mathrm{IVT}_2^{2,1}(x) = x$ is not a contraction as $\frac{d(f(x), f(y))}{d(x,y)} = 1\ \forall\, x, y \in X$ .

For the function $\mathrm{IVT}_3^{2,1}(x) = 2^k - 1$ where $x$ is $k$ bit number the contraction condition as follows:

Let $x > y$ and $x$ has higher number of bits in its representation than that of $y$.

In that case $\frac{d(\mathrm{IVT}_3^{2,1}(x), \mathrm{IVT}_3^{2,1}(y))}{d(x,y)} = \left|\frac{(2^k) - (2^l)}{(x-y)}\right|$

Now, $2^k > x, 2^{k1} > y \Rightarrow |2^k - 2^{k1}| > |x - y|$.$\frac{d(\mathrm{IVT}_3^{2,1}(x), \mathrm{IVT}_3^{2,1}(y))}{d(x,y)} > 1$. Thus$\mathrm{IVT}_3^{2,1}$ is not a contraction.

Therefore, $\mathrm{IVT}_0^{2,1}$ is the only function which satisfies the contraction condition.

It is to be noted that it is verified (by computer simulation) that no function in linearized ($A = 1, B = 0$) as well as in affine domain (non-zero B) show contraction in $p$-adic domain except the zero (trivial) function. Hence no question of global stability arises in case of non-trivial functions.

Let us now characterize the ADDS of type-I and type-II in the light of their attractors.

## *5. Characterization of ADDS:*

**Theorem 5.1** *The set of all Collatz-like ADDS of type-I is a subset of the set of all Collatz-like ADDS of type-II, with attractor* Â *where* $(\hat{A} - B)/A$ *is the attractor in subclass* $II$.

**Proof:** Let us denote the set of all ADDS of type-I and type-II as $F$ and G.

Let $\mathrm{IVT}_j^{p,1} \in F$. So $\mathrm{IVT}_j^{p,1}$ is Collatz-like function and from the definition of Collatz function it is obvious that after finitely many iterations (say $n$) $\mathrm{IVT}_j^{p,1}(x)$ will reach to an attractor and either stay there or oscillates locally around it. Let Â be the attractor of the ADDS associated with $\mathrm{IVT}_j^{p,1}$.

So for any initial value $Y_0$, $(IVT_j^{p,1})^n\,(x) = \hat{A}$

i.e. $A\mathrm{IVT}_j^{p,1}\,(A\mathrm{IVT}_j^{p,1}\,(A\mathrm{IVT}_j^{p,1}\,(A \ldots\ldots..A\mathrm{IVT}_j^{p,1}\,(Y_0) + B) + B) + B) \ldots\ldots + B) + B = \hat{A}$

The result is true for any initial value $y \in \mathrm{N}_0$.

It is possible to achieve two non-negative integers $A, B$ such that $y = AY_0 + B$, and for all $Y_0, (AY_0 + B) \in N_0$.

Substituting $Y_0$ by $Ay_0 + B$ in the above equation we get

$$A\mathrm{IVT}_j^{p,1}\,(A\mathrm{IVT}_j^{p,1}\,(A\mathrm{IVT}_j^{p,1}\,(A \ldots\ldots..A\mathrm{IVT}_j^{p,1}\,(AY_0 + B) + B) + B) + B) \ldots\ldots + B) + B = \hat{A}\ for\ all\ Y_0 \in \mathrm{N}_0.$$

$$\Rightarrow \quad \mathrm{IVT}_j^{p,1}\,(A\mathrm{IVT}_j^{p,1}\,(A\mathrm{IVT}_j^{p,1}\,(A \ldots\ldots..A\mathrm{IVT}_j^{p,1}\,(AY_0 + B) + B) + B) + B) \ldots\ldots + B) = (\hat{A} - B)/A$$

Clearly, the left hand side of the above equation equals to the nth iteration of the function $\mathrm{IVT}_j^{p,1}$ in ADDS of type-II. It is thus can be concluded that after n iterations the function $\mathrm{IVT}_j^{p,1}$ reaches a fixed attractor $(\hat{A} - B)/A$ for all $y_0 \in \mathrm{N}_0$. Therefore $\mathrm{IVT}_j^{p,1} \in$ G. So $F \subseteq G$.

*Note: The converse statement of the theorem is not true in general.*

*Justification:* Let $g \in G$.The attractor in this case is Ĝ. So after n iterations

$$g(Ag(Ag(A \ldots\ldots..Ag(Ay_0 + B) + B) + B) + B) \ldots\ldots + B) = \hat{G}\ for\ all\ y_0 \in \boldsymbol{N_0}.$$

Substituting $y_0$ by $\frac{y_0 - B}{A}$ in the above equation we get

$$g(Ag(Ag(A \ldots\ldots..Ag(y_0) + B) + B) + B) \ldots\ldots + B) = \hat{G}$$

$$\Rightarrow Ag(Ag(Ag(A \ldots\ldots..Ag(y_0) + B) + B) + B) \ldots\ldots + B) + B = A.\hat{G} + B$$

But we cannot conclude that g is Collatz like because for any $y_0 \in \mathrm{N}_0$. The quantity $(y_0 - B)/A$ may not belong to $\mathrm{N}_0$. So for all $y_0 \in \mathrm{N}_0$, g will not satisfy the above equations.So g is not Collatz-like. Hence the converse is not true, in general.

**Theorem 5.2:** *In linearized* $(A = 1, B = 0)$ $\mathrm{T}^{p,1}$ *domain,* $p^{p-2}$ *number of functions have unique steady state equilibrium point and the corresponding equilibrium point is* $0$.

**Proof:** it can be easily shown that there are $p^{p-1}$ number of collatz like functions in a p adic domain. The cases where 0 maps to $0, 1, 2, \dots, p-1$(all $p$ arguments) are equal. So among Collatz like functions 0 maps to 0 in $p^{p-2}$cases ($p^{p-1}/p$). In those cases only the existence of unique steady state point are guaranteed, and the steady state point is 0.

For rich understanding of the attractor of ADDS for different A, B (a, b) we have made computer simulations and that are resulted below in the table 1a and 1b.

| **Values of A, B** | **j value(Attractor)** | **j value (Unique steady state points)** | **Locally Stable j value (point )** | **Globally Stable j value (point)** |
|---|---|---|---|---|
| A=1,B=0 | 0(0),1(0),2(0),6(0),7(0),8(0),9(0),10(0),11(0) | 0(0),6(0), 9(0) | 0(0) | 0(0) |
| A=1,B=1 | 0(1),1(1),6(3),7(1) | 0(1),1(1) | 0(1) | 0(1) |
| A=1,B=2 | 0(2),2(2),9(2),11(2) | 0(2),2(2) | 0(2) | 0(2) |
| A=2,B=2 | 0(2),1(2),18(6) | 0(2),1(2) | 0(2) | 0(2) |
| A=2,B=0 | 0(0),1(0),2(0),3(0),4(0),18(0), 19(0),20(0) | 0(0),18(0) | 0(0) | 0(0) |
| A=2,B=1 | 0(1),2(1),3(3) | 0(1),2(1) | 0(1) | 0(1) |

*Table-1a:* Some computer simulation result in ADDS of type-I $\mathrm{Y}_{t+1} = \mathrm{A}\,\mathrm{IVT}_{j}^{p,1}(\mathrm{Y}_t) + \mathrm{B}$ in 3-adic domain.

| **Values of a, b** | **j value(Attractor)** | **j value (Unique steady state points)** | **Locally Stable j value (point )** | **Globally Stable j value (point)** |
|---|---|---|---|---|
| a=1,b=0 | 0(0),1(0),2(0),6(0),7(0),8(0),9(0),10(0),11(0) | 0(0),6(0), 9(0) | 0(0) | 0(0) |
| a=1,b=1 | 0(0),1(0),6(2),7(0) | 0(0),1(0) | 0(0) | 0(0) |
| a=1,b=2 | 0(0),2(0),9(0),11(0) | 0(0),2(0) | 0(0) | 0(0) |
| a=2,b=2 | 0(0),1(0),18(2) | 0(0),1(0) | 0(0) | 0(0) |
| a=2,b=0 | 0(0),1(0),2(0),3(0),4(0),18(0), 19(0),20(0) | 0(0),3(0)18(0) | 0(0) | 0(0) |
| a=2,b=1 | 0(0),2(0),3(1) | 0(0),2(0) | 0(0) | 0(0) |

*Table-1b:* Some computer simulation result in ADDS of type-II $y_{t+1} = IVT_j^{p,1}\,(a\,y_t + b)$ in 3-adic domain.

***Graphical Representation of Collatz like ADS:***

Here we represent a graphical view of dynamics of the ADDS of type-I (similar can be done for ADDS of type-II too). Noticeably it would very clear that for any given $Y_0$ how the ADDS behave over the iterations.

In particular in $\mathrm{T}^{3,1}$, considering $A = 1, B = 1$ four functions are Collatz like ($\mathrm{IVT}_0^{3,1}, \mathrm{IVT}_1^{3,1}, \mathrm{IVT}_6^{3,1}$ and $\mathrm{IVT}_7^{3,1}$) for the system $Y_1 = A\,IVT_j^{p,1}(Y_0) + B$.The following graphs show $Y_1 \sim Y_0$ relationship.

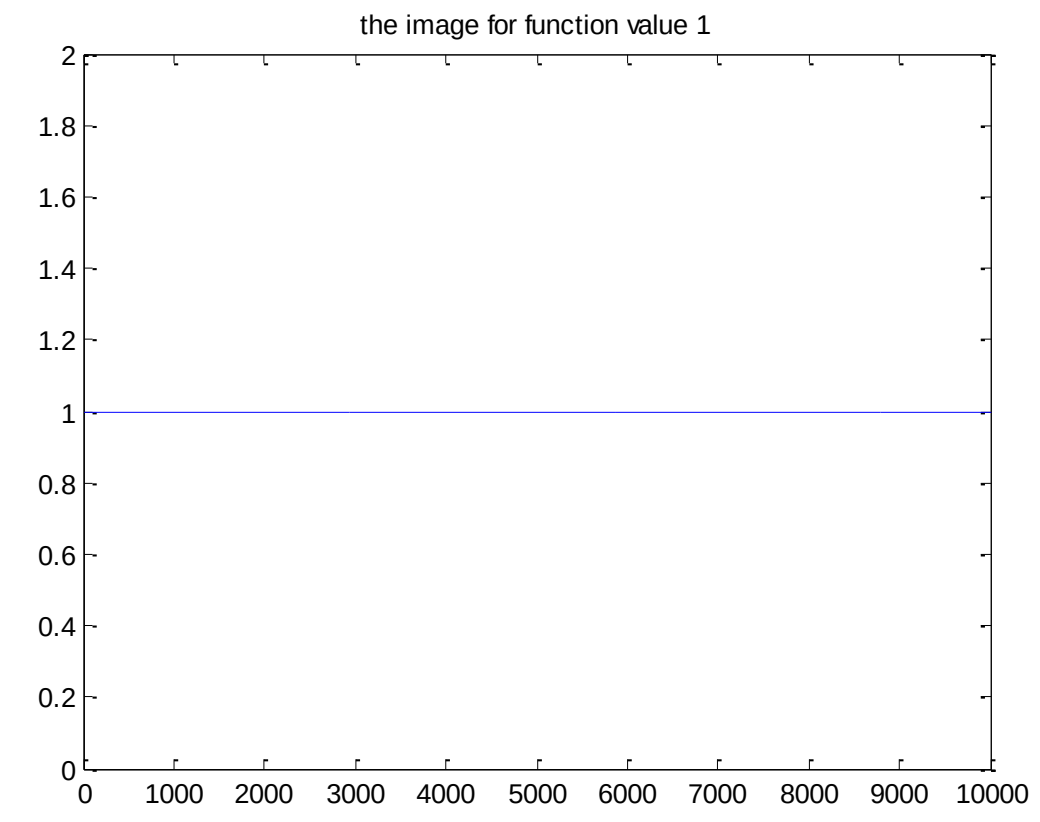

$Y_1 \sim Y_0$ of $\mathrm{IVT}_0^{3,1}$

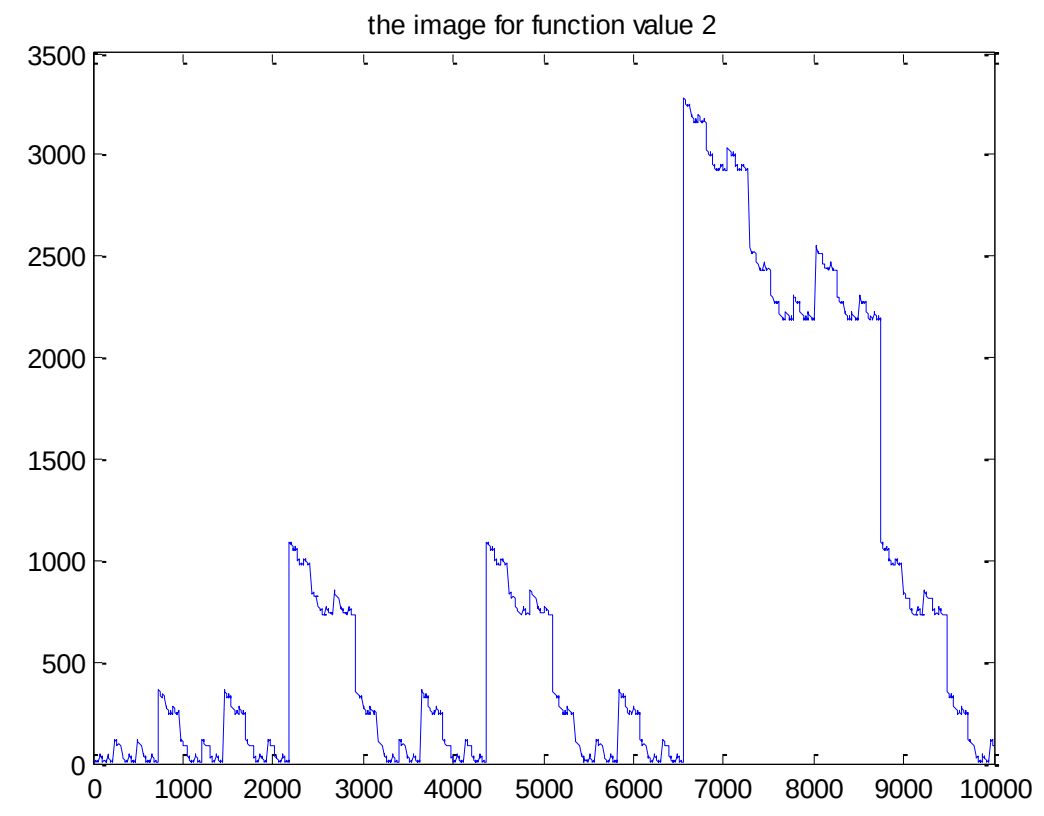

$Y_1 \sim Y_0 \sim Y_0 \mathrm{IVT}_1^{3,1}$

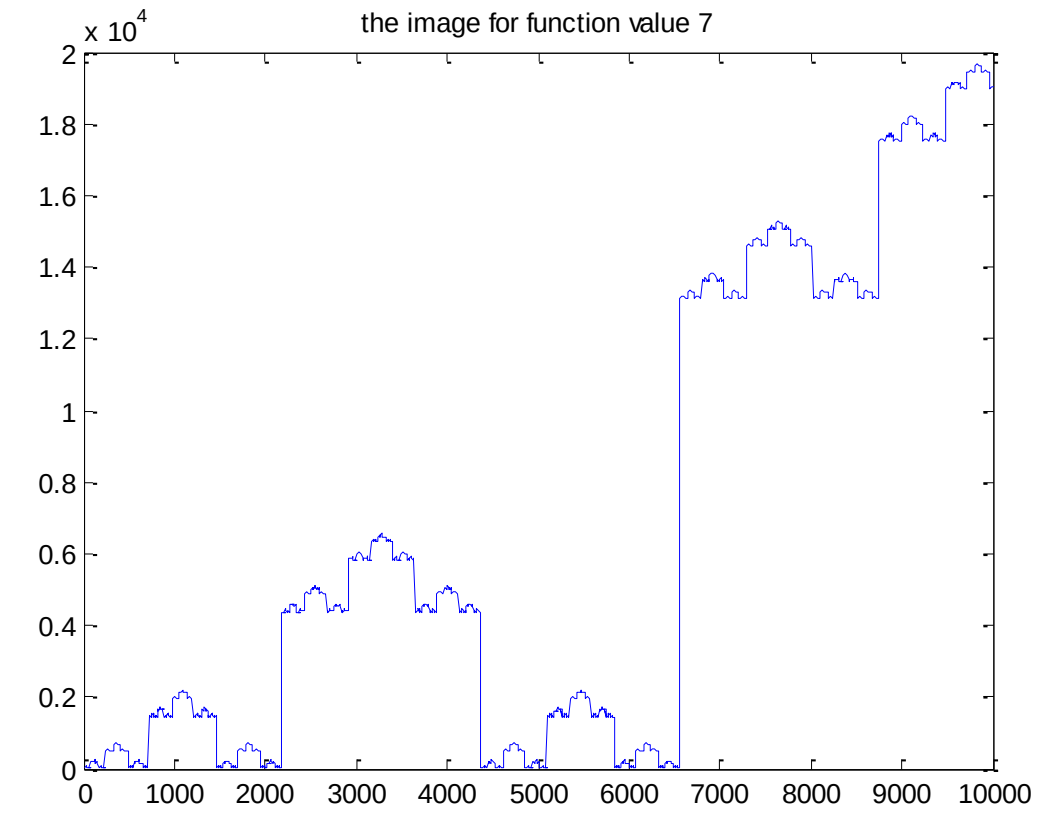

$Y_1 \sim Y_0$ of $\mathrm{IVT}_6^{3,1}$

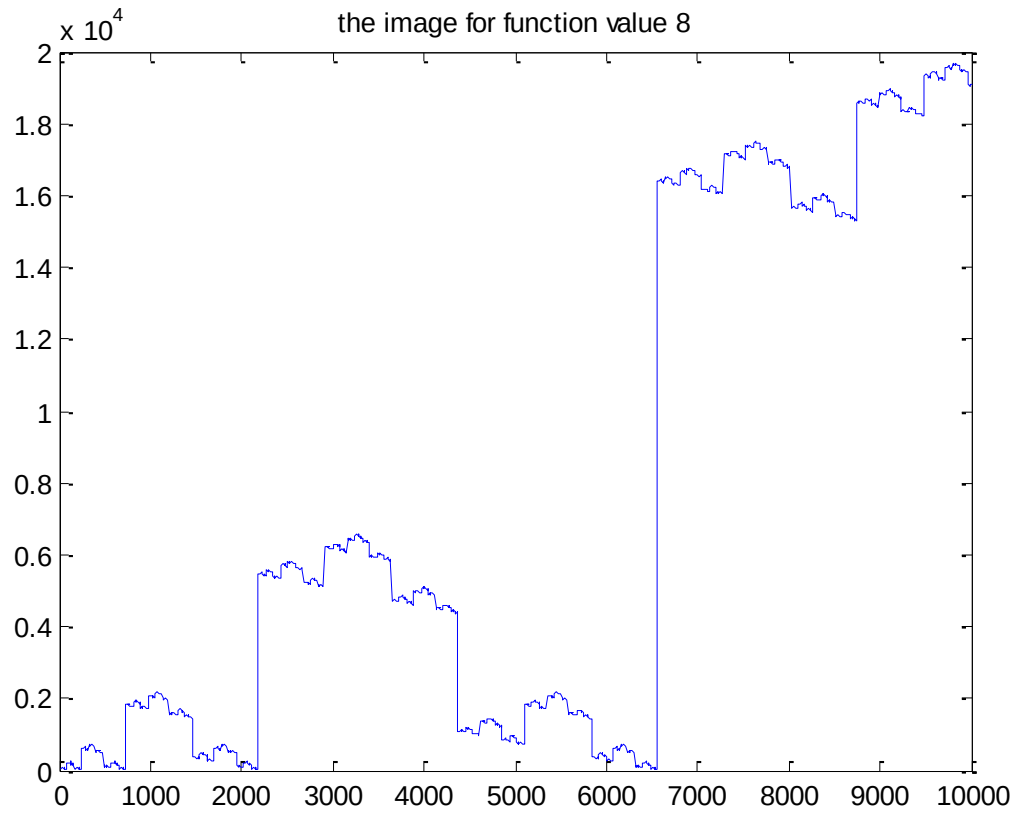

$Y_1 \sim Y_0$ of $\mathrm{IVT}_7^{3,1}$

The graphs are nowhere differentiable and self-repeating (self-similar), imply that Collatz-like ADDS forms *fractal*. The fractal dimensions of these four $IVT_j^{p,1}$s (j = 0, 1, 6, 7) are 1, 1.94006, 1.94012 and 1.94016 respectively.

Similar graphs can be obtained for the system $y_1 = IVT_j^{p,1}(A\, y_0 + B)$ and those graphs will have same pattern.

Here in general, for better understanding about non-periodicity and non-linearity of the ADDS the graph of $Y_{i+1} \sim Y_i$ of $\mathrm{IVT}_0^{3,1}$, $\mathrm{IVT}_1^{3,1}$, $\mathrm{IVT}_6^{3,1}$ and $\mathrm{IVT}_7^{3,1}$ have been sketched here.

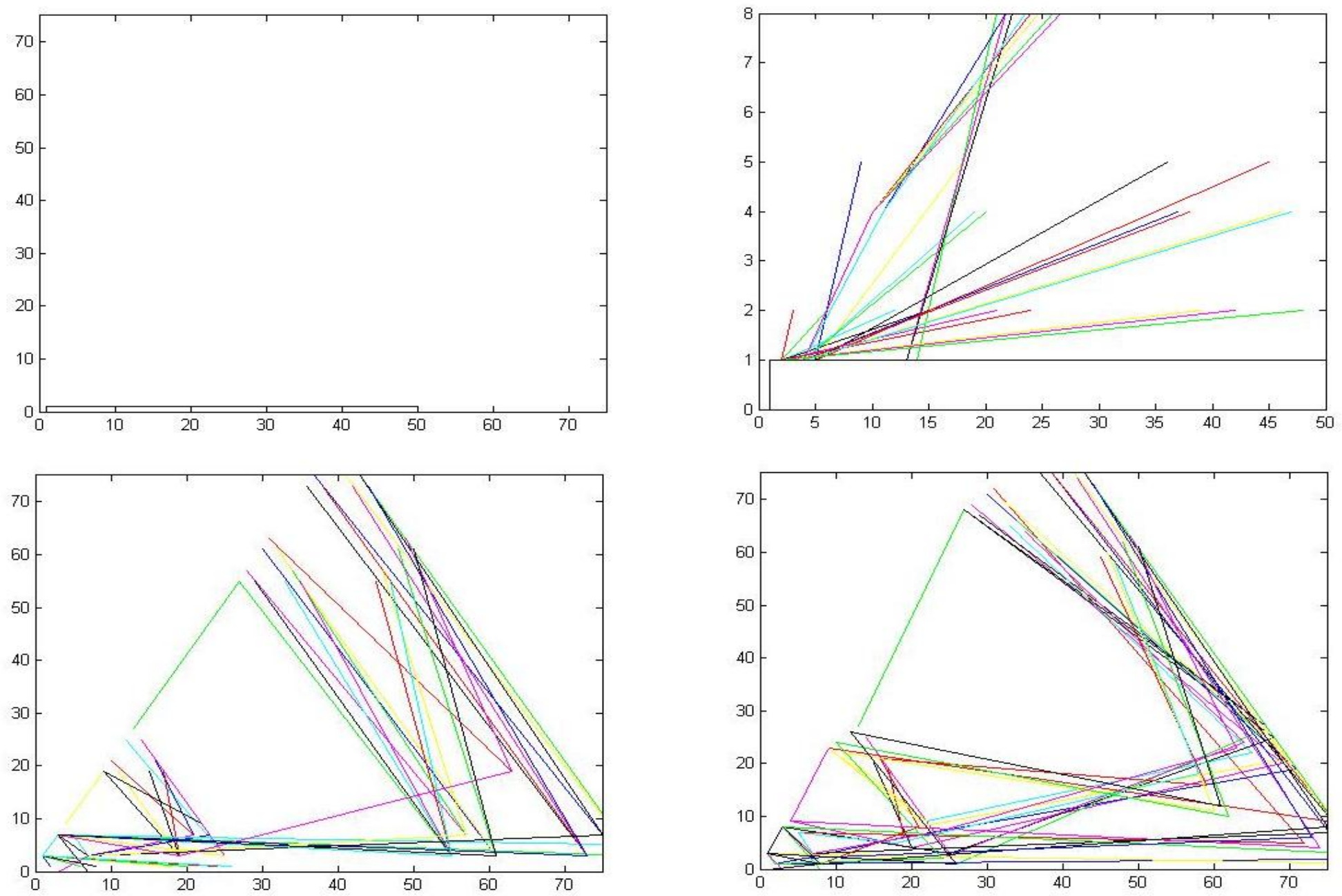

Fig. 1: *Graph of* $Y_{i+1} \sim Y_i$ *of* $IVT_0^{3,1}$, $IVT_1^{3,1}$, $IVT_6^{3,1}$ *and* $IVT_7^{3,1}$

From the Fig.1, $\mathrm{Y_{i+1}}$ vs $\mathrm{Y_i}$ graph, it is intuitively clear that the above Collatz-like IVTs are non-periodic. Also, the next result shows that every non-identity Collatz-like IVTs sequence is divergent.

**Theorem 5.3:** *There exists only one IVT in p-adic domain namely the identity map, which can be used to generate a convergent power series (with radius of convergence* $= 1$ *) in ADDS of type-I and type-II.*

**Proof:** Let the power series be $S = \sum_{n=0}^{\infty} a_n x_n$ ; $a_n = A\ IVT_j^{p,1}(n) + B$ in ADDS of type-I.

From the ratio test of convergence, the series is convergent if,

$\lim_{n\to\infty} \left|\frac{a_{n+1}x^{n+1}}{a_n x^n}\right| < 1$ i.e. $|x| < \lim_{n\to\infty} \left|\frac{a_n}{a_{n+1}}\right| = r$ , $r$ is called the radius of convergence. For the identity map $IVT_j^{p,1}(n) = n$, $\lim_{n\to\infty} \frac{|An+B|}{|An+n+B|} = 1$ where, $a_n = An + B,\quad a_{n+1} = A(n+1) + B.$

Therefore the power series is convergent for $|x| < 1$, and the radius of convergence is 1. It is verified (by simulation) that no other function exhibit the above property.

In ADDS of type-II, $a_n = IVT_j^{p,1}(an+b) = an + b,$ for identity function $\lim_{n\to\infty} \left|\frac{a_n}{a_{n+1}}\right| = \lim_{n\to\infty} \frac{|an+b|}{|an+n+b|} =$ 1. Hence the same result follows as above.

The following graphs illustrate that for the system $Y = 1.IVT_{21}^{3,1}(n) + 1$ only shows convergence property of power series. All other are leading to divergent power series. For illustrative purpose, $Y = 1.IVT_{11}^{3,1}(n) + 1$ is shown in Fig-2.

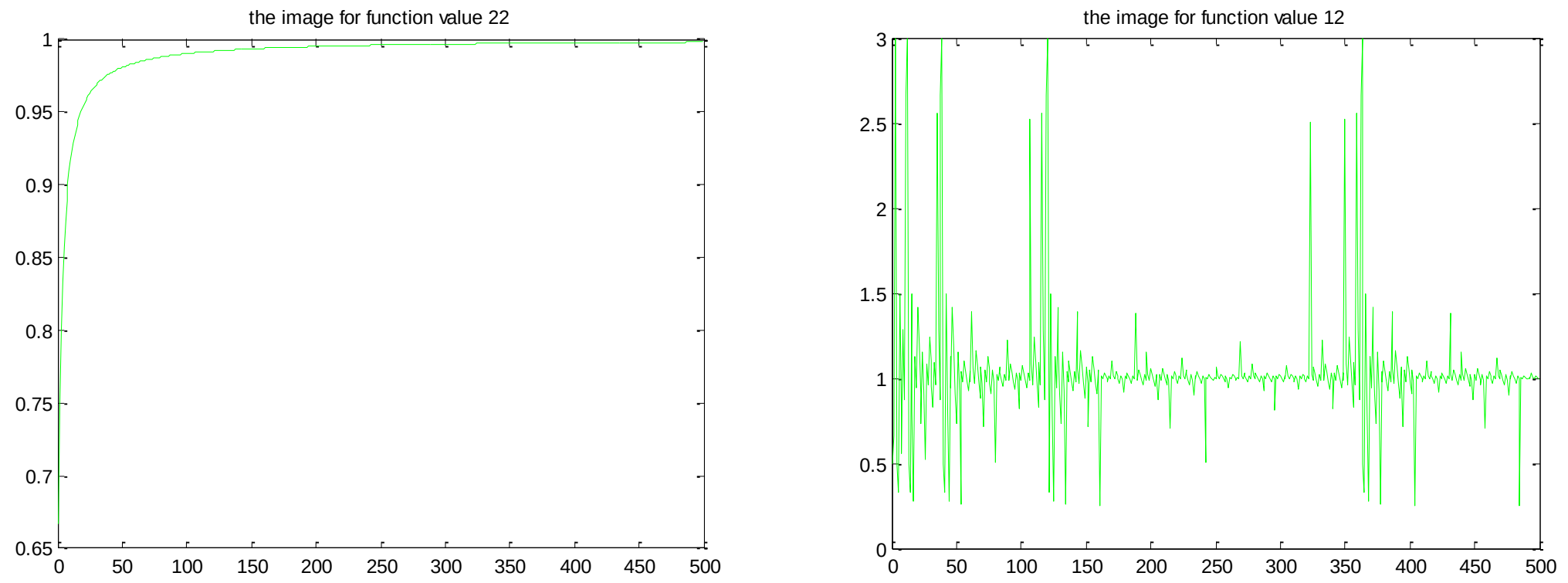

Fig 2. Graph of $\left|\frac{a_n}{a_{n+1}}\right| \sim n$ for $Y = 1.IVT_{21}^{3,1}(n) + 1$ and $Y = 1.IVT_{11}^{3,1}(n) + 1$

Let us now come to application of ADDS in distributed parallel computing environment optimal design.

## *6. Application in Distributed and Parallel Environment (DPE):*

The essence of distributed Parallel Computing (DPC) is straight forward in applied computer science [5, 6]. The central authority (often called the Super Controlling Agent) distributes a set of tasks to its immediate subordinate authorities. Further the subordinate authorities re-distribute their corresponding tasks to their subordinates and this process continues finitely. Finally, the agents residing in the lowermost level of the described architecture in parallel compute the subtasks they are given and submit it to their corresponding immediately higher authority through the path where from they got the task. This way the process of submission continues and finally the set of completed tasks are submitted to the central authority.

In particular, we have used ADDS of type-I, namely $Y_{n+1} = \mathrm{IVT}^{p,1}{}_{j}(Y_n)$.

### *6. 1 Architecture and Requirements for ODPE:*

We will use here Collatz like $\mathrm{IVT}^{p,1}{}_{j}$ in $\mathrm{T}^{p,1}$ (ADDS) in the scheduling process. It has basically a 3-layered architecture. The innermost core layer is preserved for *Super Controlling Agent* (*SCA*). This agent will only communicate with a number of *stations* who reside in the second layer. The outermost layer consists of *sub-stations*.

The main constraints which are considered here for ODPE are as the following:

I. *SCA* will communicate with minimum number of *stations*.
II. The *sub-stations* cannot directly communicate with the *SCA*.
III. There is no interaction between the *stations*.
IV. The interaction between sub-stations (i.e. hopping) is minimum.

An example of such design (the 3-layed architecture) has been made as shown in Fig. 3.

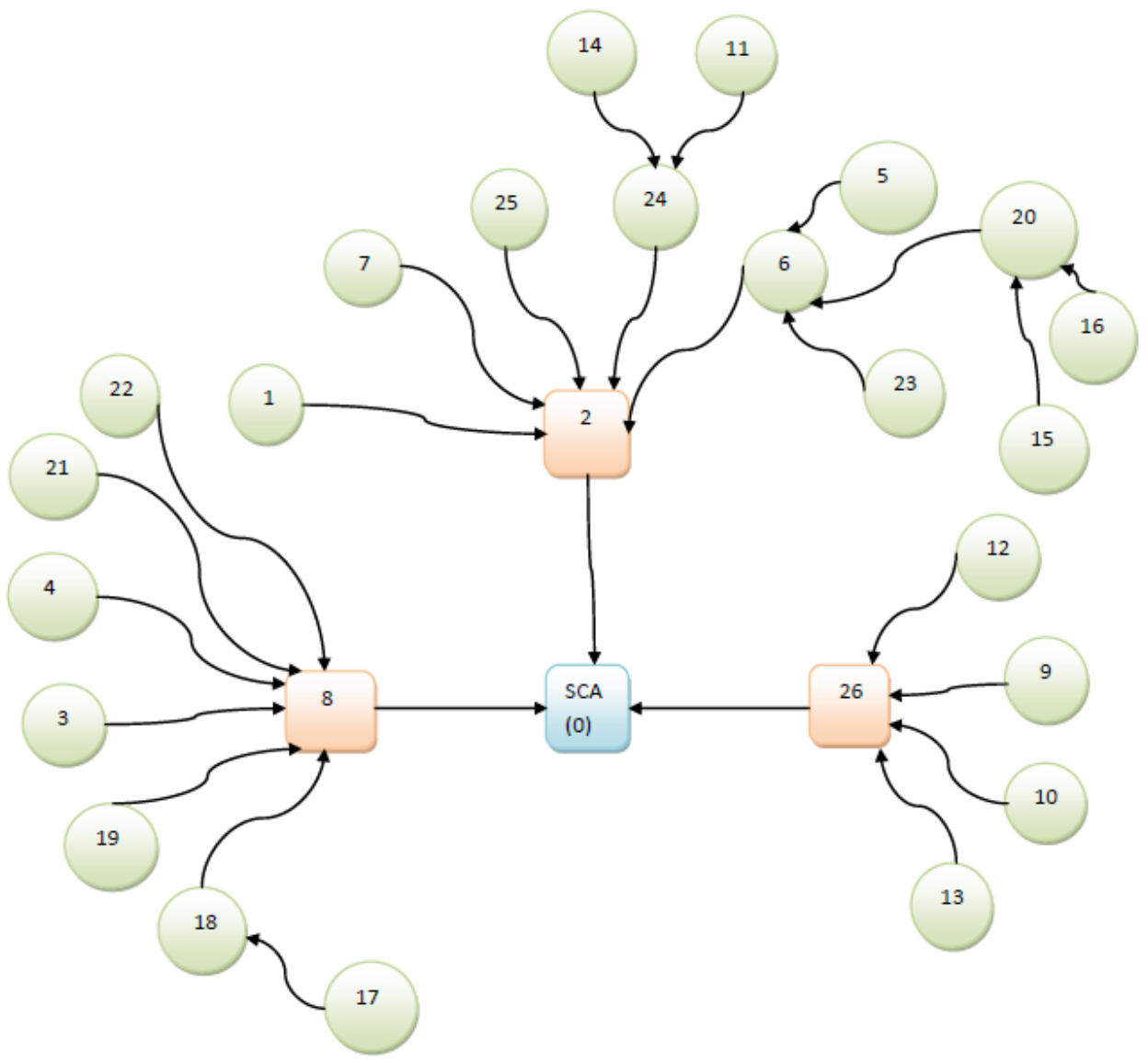

Fig.-3 Scenario of three layered architecture using $IVT_8^{3,1}$ considering 27 nodes.

Here we show the three layered architecture using $IVT_8^{3,1}$ considering total number of nodes being 27. Here 0 designate as *SCA* of the system. There are 3 *stations* (belonging to the middle layer) designated by 2 $(= 3^1 - 1)$, 8 $(= 3^{.2} - 1)$, and 26 $(= 3^3 - 1)$. The *sub-stations* (which belong in the outermost layer) either communicate with the stations directly or by a number of hops. For example *sub-stations* designated by 1, 7, and 25 communicates with 2 directly where *sub-station* 16 makes 3 hops $(16 \rightarrow 20 \rightarrow 6 \rightarrow 2)$. The *stations* communicate directly with the *SCA*.

***6. 2. Results***

The investigation of the Collatz like functions in $T^{3,1}$ is elaborated in the discussion to follow. The Collatz like in $T^{3,1}$ are $IVT_8^{3,1}$, $IVT_0^{3,1}, IVT_1^{3,1}, IVT_2^{3,1}, IVT_6^{3,1}, IVT_7^{3,1}, IVT_8^{3,1}, IVT_9^{3,1}, IVT_{10}^{3,1}$ and $IVT_{10}^{3,1}$. Since, the ADDS is linearized domain, so the SCA (attractor of the ADDS of type-I and type-II) is 0. The underlying $f_j$s are listed below for the Collatz like IVTs in $T^{3,1}$ .

| p | $f_0$ | $f_1$ | $f_2$ | $f_6$ | $f_7$ | $f_8$ | $f_9$ | $f_{10}$ | $f_{11}$ |
|---|---|---|---|---|---|---|---|---|---|
| 0 | **0** | 1 | 2 | **0** | 1 | 2 | **0** | 1 | 2 |
| 1 | **0** | **0** | **0** | 2 | 2 | 2 | **0** | 0 | 0 |
| 2 | **0** | **0** | **0** | **0** | 0 | 0 | 1 | 1 | 1 |

Table-2. *The function definitions for Collatz like functions in 3-adic system*

However, if the number of stations is to be minimum then The IVTs namely $IVT_0^{3,1}, IVT_1^{3,1}, IVT_2^{3,1}, IVT_6^{3,1}$ and $IVT_9^{3,1}$ will not satisfy, it is so because more than one entry (as shown in Table-2) for these functions map to zero. There are many natural numbers which can directly map to zero (SCA), which is not desired as per the constraint (I) stated above.

Excluding the functions defined above we are left with 4 IVTs namely $\mathrm{IVT}_7^{3,1}, \mathrm{IVT}_8^{3,1}, \mathrm{IVT}_{10}^{3,1}, \mathrm{IVT}_{11}^{3,1}$. These IVTs are analyzed by classifying into two sets, $\{\mathrm{IVT}_7^{3,1}, \mathrm{IVT}_8^{3,1}\}$ and $\{\mathrm{IVT}_{10}^{3,1}, \mathrm{IVT}_{11}^{3,1}\}$.

For $IVT_7^{3,1}, IVT_8^{3,1}$ the stations that will be ultimately attracted by the SCA in 1 iteration for an n digit 3-adic system are given as,

$$2.3^0 + 2.3^1 + \cdots n\ terms = 2.\frac{3^n - 1}{3 - 1} = 3^n - 1$$

The number of stations must be less as only $3^n - 1$ will behave as stations.

For $IVT_{10}^{3,1}, IVT_{11}^{3,1}$ we have the number of stations with SCA as an attractor given as follows

$$1.3^0 + 1.3^1 + \cdots n\ terms = 1.\frac{3^n - 1}{3 - 1} = \frac{3^n - 1}{2}$$

For the latter case, the number of stations will be more than $IVT_7^{3,1}, IVT_8^{3,1}$ which is clear from the expressions of stations in both cases.

Another fact of importance is the *number of hopping*. It is desired that the number of hopping is minimum. In, $IVT_8^{3,1}$ both 0, 1 map to 2 while in $IVT_7^{3,1}$ it is not. Intuitively it can be told that in $IVT_8^{3,1}$ hopping is minimum as the number of steps to reach SCA (0) will be less. To clarify the understanding a parameter namely *Average Hopping* is defined.

$$\text{Avg Hopping} = \frac{(\text{Total number of hops})}{(\text{The natural number upto which hopping is checked})}$$

The proposition is checked with Matlab Codes.

The following Fig-3 shows hopping ~ argument (natural number) relationship (the argument is taken upto 100) and infer that average hopping is less in case of $IVT_8^{3,1}$.

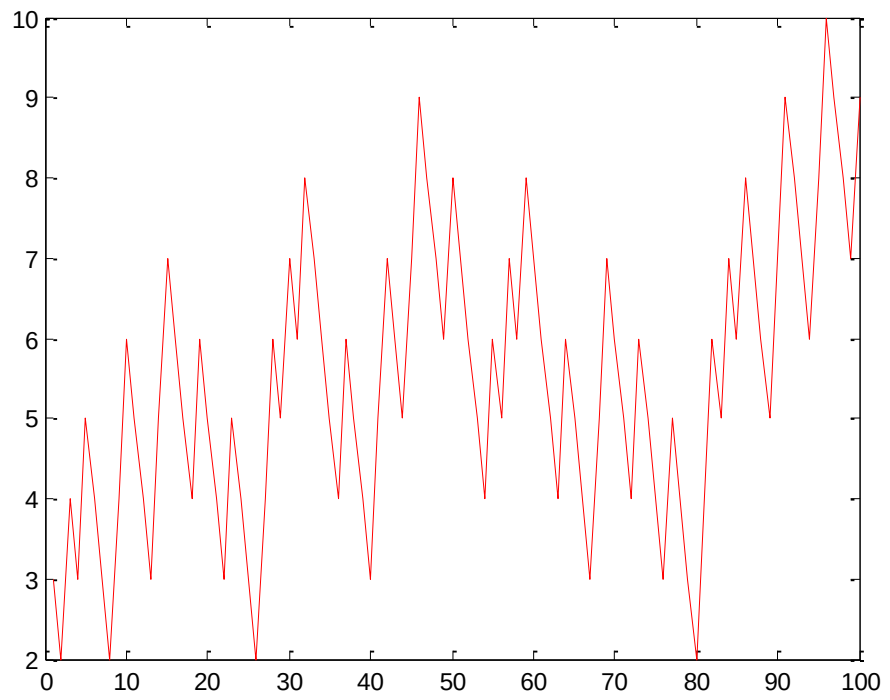

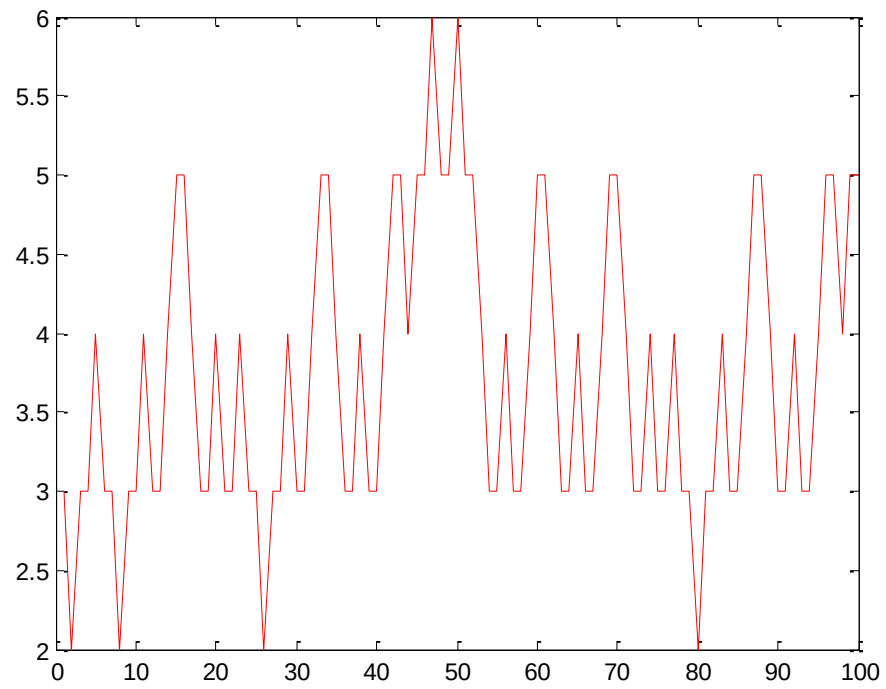

Fig. 4: $f_7, avg\ hop =$ 6.46 $\qquad$ $f_8, avg\ hop$ =4.73

So, considering all the facts $IVT_8^{3,1}$ is the most desirable IVTs in $\mathrm{T}^{\mathrm{p},1}$ for scheduling.

In p=2, $IVT_1^{2,1}$ is the most desirable (only practically feasible) as the only other Collatz like function is the trivial function.

Another issue may be raised about the capacity of the system, i.e. if ideally N number of nodes are considered, no intermediate mapping will exceed N.

This problem can be resolved by simply avoiding the nodes which will map to an unnecessary large number which exceeds the capacity of the system. For an example, in $T^{p,1}$ every natural number upto 80 maps to numbers less than 80 but 81 maps to 242. The policy is not to assign any substation by 80 or simply avoid it.

**Generalized Corollary 7.2:** *In* $T^{p,1}$, $IVT^{p,1}_{p^{p-1}-1}$ *will have the best scheduling possibility and stations will be designated by* $p^n - 1$.

Proof of the corollary is straightforward from the above demonstration.

These Collatz like IVTs have a great potentiality in optimal design of DPE.

## *7. Conclusion and Future Research Effort:*

In this paper, a primary study on Collatz like ADDS of type-I and type-II have been presented and using these ADDS an ODPE have been designed. The study of non-linear, multi-dimensional discrete dynamical systems and theory characterization in the light of IVTs would be our future endeavors. In practice, the constraints which are so far we have addressed in designing an ODPE are not sufficient. So our future research exertion, we will addressed more feasible and practical constraint depending upon nature of the DPE and where the mathematics of ADDS can be utilized properly.

***Acknowledgement:*** The authors would like to thank to *Prof. B. S. Daya Sagar* for his valuable suggestions and comments.